\documentclass[aps,prl,twocolumn,superscriptaddress,showpacs,reprint]{revtex4-1}

\usepackage[dvips]{graphicx}
\usepackage{dcolumn}
\usepackage{bm}
\usepackage[reqno, intlimits,]{amsmath}
\usepackage[latin1]{inputenc}
\usepackage{textcomp}
\usepackage{color}

\def\nablav{\mbox{\boldmath$\nabla$}}

\def\rot{\nablav\times}
\def\xv{\hat{\bf x}}
\def\yv{\hat{\bf y}}
\def\zv{\hat{\bf z}}

\def\nv{\hat{\bf n}}

\def\bea{\begin{eqnarray}}
\def\eea{\end{eqnarray}}

\begin{document}

\title
{Laser-Driven Rayleigh-Taylor Instability:\\Plasmonics Effects and Three-Dimensional Structures}

\author{A.~Sgattoni}\email[ ]{andrea.sgattoni@polimi.it}
\affiliation{Dipartimento di Energia, Politecnico di Milano, Milano, Italy}
\affiliation{Istituto Nazionale di Ottica, Consiglio Nazionale delle Ricerche, research unit ``Adriano Gozzini'', Pisa, Italy}
\author{S.~Sinigardi}
\affiliation{Istituto Nazionale di Ottica, Consiglio Nazionale delle Ricerche, research unit ``Adriano Gozzini'', Pisa, Italy}
\affiliation{Dipartimento di Fisica e Astronomia, Universit\`a di Bologna, via Irnerio 46, 40126 Bologna, Italy}
\affiliation{INFN sezione di Bologna, viale Berti Pichat 6/2, 40127 Bologna, Italy}
\author{L.~Fedeli}
\author{F.~Pegoraro}
\author{A.~Macchi}\email[ ]{andrea.macchi@ino.it}
\affiliation{Istituto Nazionale di Ottica, Consiglio Nazionale delle Ricerche, research unit ``Adriano Gozzini'', Pisa, Italy}
\affiliation{Dipartimento di Fisica ``Enrico Fermi'', Universit\`a di Pisa, Largo Bruno Pontecorvo 3, I-56127 Pisa, Italy}

\date{\today}

\begin{abstract}
The acceleration of dense targets driven by the radiation pressure of high-intensity lasers leads to a Rayleigh-Taylor instability (RTI) with rippling of the interaction surface. Using a simple model it is shown that the self-consistent modulation of the radiation pressure caused by a sinusoidal rippling affects substantially the wavevector spectrum of the RTI depending on the laser polarization. The plasmonic enhancement of the local field when the rippling period is close to a laser wavelength sets the dominant RTI scale. The nonlinear evolution is investigated by three dimensional simulations, which show the formation of stable structures with ``wallpaper'' symmetry. 
\end{abstract}



\maketitle
The Rayleigh-Taylor instability (RTI) is the classical process occurring when a heavy fluid stands over a lighter one in hydrodynamics or, equivalently, when a light fluid accelerates a heavier one. The latter case is of crucial importance in Inertial Confinement Fusion \cite{atzeni-book} and in the astrophysical context, as exemplified in a spectacular way by the Hubble Space Telescope images of the Crab Nebula \footnote{\protect\url{http://hubblesite.org/newscenter/archive/releases/2005/37/}}. A peculiar example of the RTI arises in the context of ultraintense laser-plasma interactions where the radiation pressure of the laser pulse is large  enough to drive a strong acceleration of a dense plasma target. Surface rippling attributed to RTI-like phenomena has been observed in simulations since early investigations of the ultraintense regime \cite{wilksPRL92} and in several studies devoted to the concept of radiation pressure acceleration of thin targets \cite{pegoraroPRL07,robinsonNJP08,klimoPRSTAB08,*chenPoP08,chenPoP11,khudikPoP14}, i.e. the ``light sail'' scheme which is being extensively studied experimentally \cite{henigPRL09b,*dollarPRL12,*kimPRL13,*aurandNJP13,*steinkePRSTAB13,*karPRL12,palmerPRL12} as one of the most promising approaches to laser-plasma acceleration of ions \cite{daidoRPP2012,*macchiRMP13,*fernandezNF14} especially at intensities beyond $10^{23}~{\mbox{W cm}^{-2}}$ (foreseen with next generation laser facilities), i.e. in the regime where the ions become relativistic and high energy gain is predicted \cite{esirkepovPRL04,tamburiniPRE12,macchiPPCF13}. The RTI may cause early breakthrough of the laser pulse through the thin foil target, leading to inefficient acceleration. Some experimental evidence of radiation pressure-driven RTI in thin targets has been reported \cite{palmerPRL12}. 

Analytical modeling of the laser-driven RTI of a thin foil in the ultraintense regime \cite{pegoraroPRL07,khudikPoP14} predicts the instability growth rate $\gamma_{\mbox{\tiny RT}}$ to increase monotonically with the wavevector $q$, similarly to the classic result for the hydrodynamic instability $\gamma_{\mbox{\tiny RT}}=(gq)^{1/2}$ where $g$ is the acceleration, thus apparently favoring the generation of small scales. However, simulations show that the size of the structures generated by the instability is finite and close to the laser wavelength \cite{pegoraroPRL07,chenPoP11,khudikPoP14}. To explain such a feature, in this Letter we consider the effect of the transverse modulation of the radiation pressure caused by the rippling of the laser-plasma interface. By studying the reflection of a plane monochromatic wave by a shallow sinusoidal grating we show that when the laser polarization is not parallel to the grating grooves the local radiation pressure can be significantly enhanced in the valleys of the grating, particularly when the grating period is close to the laser wavelength. We calculate the modified linear growth rate and the unstable wavevector spectrum which is found to depend on the laser polarization. Our analysis is supported by particle-in-cell (PIC) simulations in two (2D) and three dimensions (3D). For circular polarization the observed hexagonal shape of RTI-generated structures resembles that predicted on the general basis of symmetry considerations. 

\begin{figure}
\centering
\includegraphics[width=0.5\textwidth]{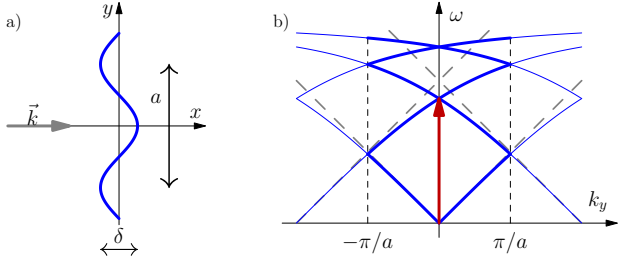}
\caption{(color online) a) the geometry. b): phase matching with surface waves at normal incidence. \label{geometry}\label{matching}}
\end{figure}

The general problem of the reflection of an electromagnetic (EM) wave from surfaces having various modulations and arbitrary refraction index and of related phenomena, such as local field enhancement and excitation of surface waves and plasmonic modes, has a long history (see e.g. \cite{fanoJOSA41,*petitNRO75,*toigoPRB75,*chandezonJOSA82,*weberPRB83}) and can be solved exactly in many cases of interest. However, for our aim it will be sufficient to consider normal incidence on perfectly reflective and shallow gratings, whose depth is small with respect to the laser wavelength, and to use a perturbative approach. We consider a plane monochromatic wave of frequency $\omega$ impinging along the $x$ direction on a perfect mirror filling the $x>x_m(y)$ region where $x_m=({\delta}/{2})\cos qy$ describes the sinusoidal rippling of the mirror surface (see Fig.\ref{geometry}~a), with $q=2\pi/a$ being $a$ the period of the ripple and $\delta$ the peak-to-valley depth. Note that $y=0$ corresponds to a valley of the rippling. Our aim is to evaluate the EM field at all  points on the surface, via a perturbative approach in the small parameter $k\delta$ where $k=\omega/c$. 

Let the electric field of the incident wave to be ${\bf E}_i=(E_{ip}\yv+E_{is}\zv)e^{ikx-i\omega t}$ where $E_{ip}=E_0$ and $E_{is}=0$ corresponds to $P$-polarization, $E_{ip}=0$ and $E_{is}=E_0$ to $S$-polarization, and $E_{ip}=iE_{is}=E_0/\sqrt{2}$ to circular ($C$) polarization, respectively.
From the general solution of Helmholtz's equation, taking account the symmetry and periodicity of the system, the components of the electric and magnetic field along $z$ (i.e. parallel to the grating grooves) may be written as (we omit everywhere the harmonic temporal dependence $\sim \exp(-i\omega t)$)
\bea
E_z &\simeq& E_{is} e^{ikx}-E_{rs}e^{-ikx}+\sum_{\ell=1}^{\infty}E_{\ell}e^{\kappa_{\ell} x}\cos(\ell qy) \; , 
\label{eq:Ez}\\
B_z &\simeq& E_{ip}e^{ikx}+E_{rp}e^{-ikx}+\sum_{\ell=1}^{\infty}B_{\ell}e^{\kappa_{\ell} x}\cos(\ell qy) \; , 
\label{eq:Bz}
\eea
where $\kappa^2_{\ell}=(l^2q^2-k^2)$. Other components are simply obtained from ${\bf B}=-i\rot{\bf E}/k$ and ${\bf E}=i\rot{\bf B}/k$.
Modes with $\ell q>k$ and $\kappa_{\ell}$ real are evanescent modes, while $\ell q<k$ and imaginary $\kappa_{\ell}$ correspond to propagating waves scattered at an angle $\alpha$ with respect to the normal direction such that $\tan\alpha=\ell q/|\kappa_{\ell}|$. 
For a shallow modulation we assume that $E_{\ell}$ is of order ${\cal O}(k^l\delta^l)$ and we truncate the expansion (\ref{eq:Ez}) up to $\ell=1$. In such case, the terms of order ${\cal O}(k\delta)$ are only evanescent for $q>k$ (grating with sub-wavelength period) and only propagating for $q<k$. 
The boundary conditions at the surface are
$E_z(x=x_m(y),y)=0$,  $({\bf B}\cdot\nv)(x=x_m(y),y)=0$, $({\bf E}\times\nv)(x=x_m(y),y)=0$ where $\nv=(-\xv+ x'_m\yv)(1+x^{'2}_m)^{1/2}\simeq-\left(\xv+\yv(q\delta/2)\sin qy\right)+{\cal O}(k^2\delta^2)$ is the unit vector normal to the surface.
We thus obtain $E_{rs}=E_{is}$, $E_{rp}=E_{ip}$, $E_{1x}=-i({qk\delta}/{\kappa_1})E_{ip}$, $E_{1y}=-ik\delta E_{ip}$, $E_{1z}=-ik\delta E_{is}$, $B_{1x}={q\delta}E_{is}$, $B_{1y} =\kappa_1\delta E_{is}$, $B_{1z}=({k^2\delta}/{\kappa_1})E_{ip}$.

For $P$-polarization, in the limit $q\to k$ we have $|\kappa_1|\to 0$ and thus $E_{1y}$ and $B_{1z}$ diverge. This is due to the excitation of a resonant, standing surface wave (SW) in the periodic medium: in fact, because of the folding of the SW dispersion relation in the Brillouin zone $|k|<\pi/a=q/2$ (see Fig.\ref{matching} b)), there is an intersection between the dispersion curves of the EM wave at normal incidence ($k_y=0$) and of the SW: for a collisionless plasma and a sufficiently shallow grating, $k_y=(\omega/c)(\omega_p^2/\omega^2-1)^{1/2}(\omega_p^2/\omega^2-2)^{1/2}\to \omega/c$ in the perfect mirror limit $\omega_p/\omega\to\infty$, with $\omega$ the plasma frequency; the latter case is equivalent to $q=k$, i.e. to a grating period equal to the laser wavelength.
The EM wave is able to excite the SW because of the modulation, so that the electric field can drive surface charge and current densities also at normal incidence. The inversion symmetry imposes that a superposition of $+q$ and $-q$ modes, i.e. a standing wave, is excited. 

The flow of the EM momentum through the mirror surface is given by ${\bf P}={\sf T}\cdot \nv$ where where ${\sf T}_{\alpha\beta}=(1/8\pi)\left[\mbox{Re}(E_{\alpha}E^*_{\beta}+B_{\alpha}B^*_{\beta})-\frac{1}{2}(|{\bf E}|^2+|{\bf B}|^2)\delta_{\alpha\beta}\right]_{x=x_m(y)}$ is Maxwell's stress tensor evaluated at the surface and averaged over an oscillation period. For $S$- and $P$-polarization we obtain for the $P_x$ component, up to order ${\cal O}(k\delta)$
\bea
P_x \simeq \frac{E_0^2}{4\pi}\left\{
\begin{array}{ll}
\mbox{Re}(1-\kappa_1\delta \cos qy) & (S) \; , 
\\
\mbox{Re}\left(1+\dfrac{k}{\kappa_1}k\delta \cos qy\right) & (P) \; ,
\end{array}\right.
\eea
while $P_y \simeq ({E_0^2}/{8\pi})q\delta\sin qy$ for both polarizations.
We thus see that to first order there is no transverse modulation in $P_x$ when $q<k$ since in this case $\kappa_1$ is an imaginary number. The modulation, to order ${\cal O}(k\delta)$ occurs only for $q>k$ and it is due to the field enhancement associated to the evanescent modes.

From now on we assume $q>k$ and consider the effects of the modulated radiation pressure on the RTI of a thin foil. Noting that ${E_0^2}/{4\pi}=2I/c\equiv P_0$ the local pressure due to EM momentum flow normal to the surface is 
\bea
P_{\perp}=-{\bf P}\cdot{\nv}\simeq P_0(1+K(q)\delta \cos qy) \; ,
\label{eq:Pmod}
\eea
where $K(q)=-\kappa_1=-(q^2-k^2)^{1/2}$ for $S$-polarization, $K(q)=(k^2q/\kappa_1)$ for $P$-polarization and $K(q)=(2k^2-q^2)/2\kappa_1$ for $C$-polarization. 
Eq.(\ref{eq:Pmod}) implies that when a surface rippling occurs the radiation pressure will be modulated in the transverse direction with a different phase depending on the polarization: for $P$- and $C$-polarization $P_{\perp}$ locally higher in the valleys and lower at the peaks, thus enforcing the growth of the modulation, while the opposite holds for $S$-polarization.

\begin{figure}
\includegraphics[width=0.5\textwidth]{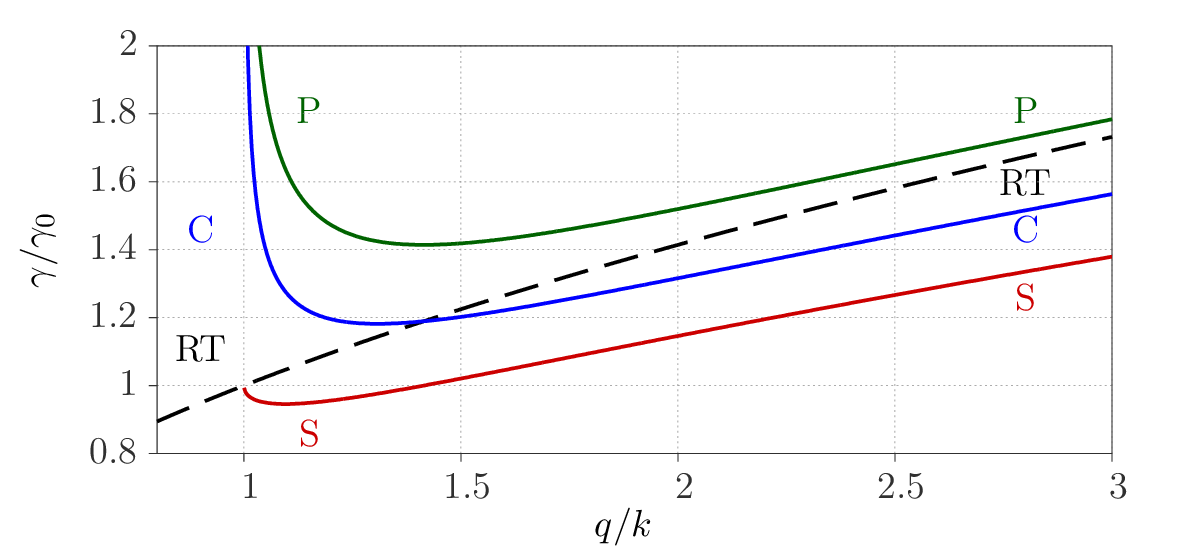}
\caption{(color online) RTI growth rate of a thin foil for $S$- (S), $P$- (P) and circular (C) polarization. The dashed curve $\gamma_{RT}=({P_0q/\sigma})^{1/2}$ gives the rate for the standard RTI of a ``flat'' thin foil. The parameter $\gamma_0=({P_0k/\sigma})^{1/2}$. \label{fig:rate}}
\end{figure}

To analyze this effect further we use the model of Ott \cite{ottPRL72} for the RTI of a thin foil driven by a pressure difference between the two sides. A similar extension of this model has been used in Ref.\cite{pegoraroPRL07} to study the relativistic regime of the instability. Here for simplicity we restrict to the non-relativistic case. We consider a thin foil of surface density $\sigma$, initially plane and placed at the position $x=0$, with a pressure $P$ on the $x<0$ side. Using Lagrangian coordinates ${\bf r}={\bf r}({\bf r}_0,t)$, 
the equation of motion for an infinitesimal fluid element of length $d{\bf r}={\bf r}(y_0+dy_0,t)-{\bf r}(y_0,t)$ and mass $dm=\sigma dy_0$ is
\bea
\partial_t^2{\bf r}=({P}/{\sigma})\left(\xv{\partial_0 y}-\yv{\partial_0 x}\right) \; ,
\eea
where $\partial_0\equiv\partial/\partial y_0$. We look for an approximate solution in the form
\bea
x(y_0,t) &\simeq& \xi_0(t)+\frac{1}{2}\xi_x(t)e^{iqy_0}+\mbox{c.c.} \; , \\
y(y_0,t) &\simeq& y_0+\frac{1}{2}\xi_y(t)e^{iqy_0}+\mbox{c.c.} \; . 
\eea
As noted in Ref.\cite{ottPRL72} such solution is not generally sinusoidal in Eulerian variables but becomes so for small perturbations ($q|\xi_i|\ll 1$), which is consistent with our calculation of the pressure modulation. Thus we substitute $P$ with $P_0(1+K(q)\xi_x)$. To lowest order the equation of motion yields $\xi_0(t)=(P_0/\sigma)t^2/2$ which describes the motion of the ``flat'' foil. To next order the equations for $\xi_x$ and $\xi_y$ are
\bea
\partial_t^2\xi_x&=&({P_0}/{\sigma})\left(K(q)\xi_x+{\partial_0\xi_y}\right) \; ,\\
\partial_t^2\xi_y&=&-({P_0}/{\sigma}){\partial_0\xi_x} \; ,
\eea 
which have solutions of the form $\xi_x \sim e^{\gamma t}$ and an unstable root ($\gamma$ real and positive) given by
\bea
\gamma=\left({P_0}/{\sigma}\right)^{1/2}\left[\left(q^2+\frac{K^2(q)}{4}\right)^{1/2}+\frac{K(q)}{2}\right]^{1/2} \; .
\eea
The growth rate $\gamma$ is shown in Fig.\ref{fig:rate} for the three polarizations and compared with the standard result $\gamma_{\mbox{\tiny RT}}=({P_0q}/{\sigma})^{1/2}$. 

Results similar to Fig.\ref{fig:rate} are expected for thick targets because the dominant effect is the plasmonic enhancement of the field at the surface, which only depends on the period of the surface rippling. The dependence on the polarization needs a careful discussion because for high intensity the laser-plasma coupling is highly sensitive to the laser polarization. For instance, in the case of $P$-polarization strong electron heating occurs and the surface rippling may be ``washed out'' by the quiver motion of electrons. We expect the above theory to be mostly appropriate for circular polarization (the preferred option for radiation pressure acceleration), for which electron heating is strongly reduced and no anisotropy in the transverse plane is generated. As far as effects of higher order in $k\delta$ are of concern, we speculate that when $k\delta \sim 1$ the field may be screened into the grating valleys if $q>2k$, similarly to what happens in a waveguide, thus reducing the RTI growth for such high-$q$ modes. 

\begin{figure}[t!]
\centering
\includegraphics[width=0.5\textwidth]{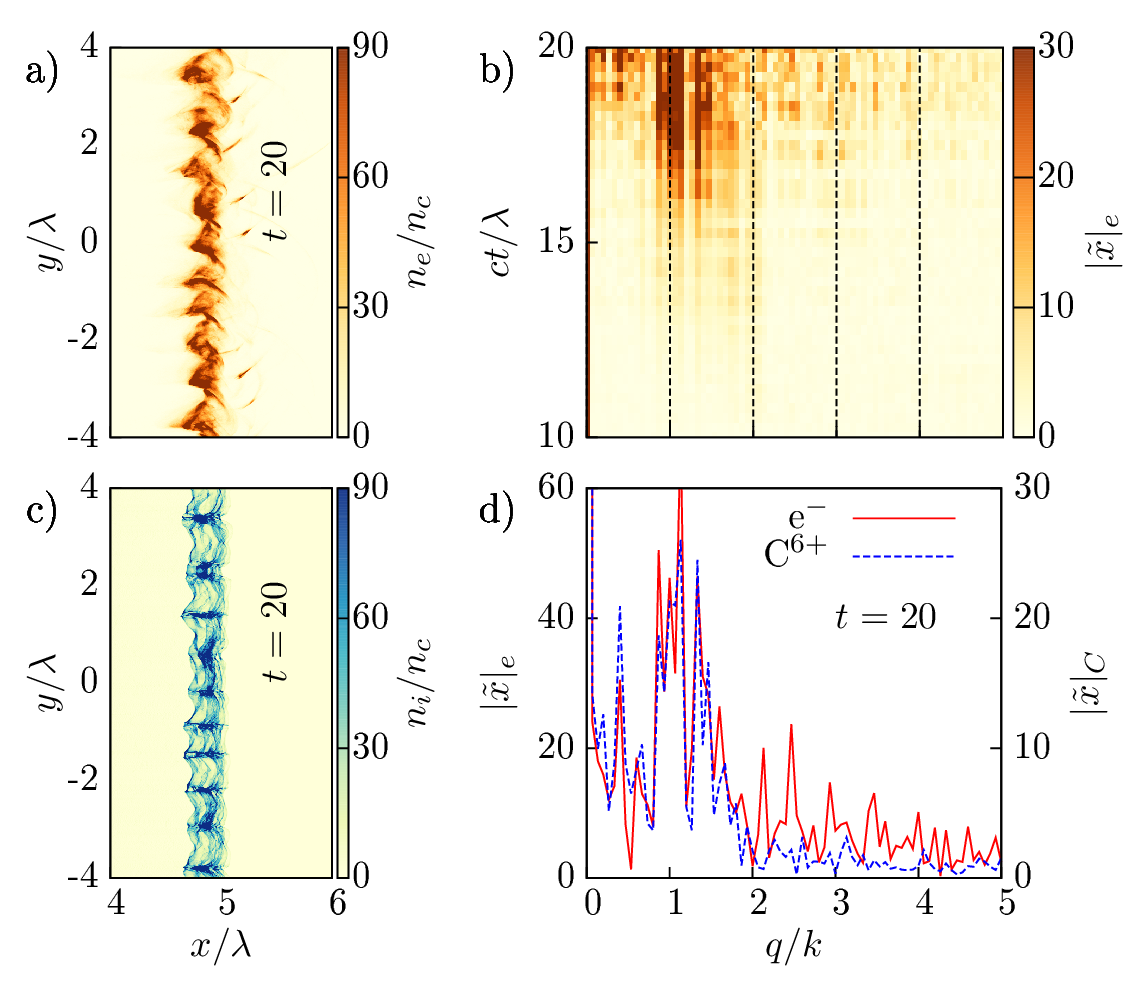}
\caption{(color online) Analysis of the transverse modes in 2D plane wave simulations. a) and c): charge density ($t=20\lambda/c$) of electrons and ions. For each time-step the longitudinal position $x=x(y,t)$ of the vacuum-plasma interface was reconstructed as a function of the transverse coordinate $y$. b): temporal evolution of the Fourier transform $\tilde{x}(q,t)$ for electrons. d): comparison of $\tilde{x}$ for electrons and ions at $t=20$.
\label{spectra}}

\end{figure}

In order to test the analytical model we performed 2D PIC simulations using the {\tt PICCANTE} open source code \footnote{\protect\url{https://github.com/ALaDyn/piccante}. DOI:10.5281/zenodo.10587}.
We considered a circularly polarized plane wave, irradiating a thin overdense ``carbon'' plasma slab (ions $Z/A=0.5$).
A periodic rippling of the foil is observed and the continuos translational symmetry is quickly broken. In  Fig.~\ref{spectra} we show the results for a simulation performed with normalized wave amplitude $a_0=(I/2m_ec^3n_c)^{1/2}=66$, target thickness $d=0.58\lambda$ and density $n_e=37n_c$ (where $n_c$ is the cut-off density). The transverse size of the simulation box was $L_y=15\lambda$ and the spatial resolution $\Delta x =\Delta y= \lambda/204$.
A Fourier analysis of the target profile shows that the dominant modes during the onset of the instability are in the range $q\approx[0.8:2]k$ (Fig.\ref{spectra}b). As expected, electrons and ions show a very similar behaviour see Fig.\ref{spectra}a,c,d. In a very wide range of simulation parameters, we consistently observe the same behaviour and a rather sharp cut-off for the modes $q>2k$ confirming a waveguide-like screening.
In contrast, for linear $P$-polarization the RTI is quenched as anticipated above. 

\begin{figure}[t]
\centering
\includegraphics[width=0.5\textwidth]{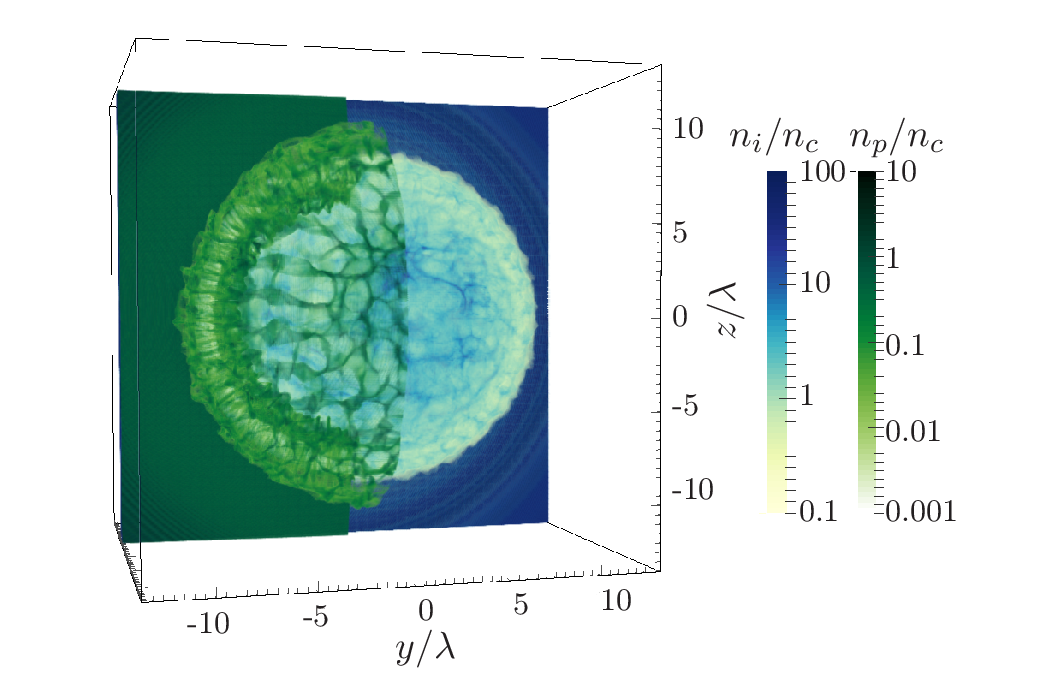}
\caption{(color online) Snapshot at $t=30T$ showing the densities of Carbon ions (for $|y|\leq 15$ and $|z|\leq 15]$, blue tones) and of protons (for $-15 \leq y\leq 0$ and $|z|\leq 15$, green tones). 
\label{fig_3D}}
\end{figure}

Large scale 3D simulations were performed with the PIC code \texttt{ALaDyn} \cite{benedettiIEEE08,*londrilloNIMA10} in the ultra-relativistic regime, for parameters close to Ref.\cite{tamburiniPRE12} and of relevance for radiation pressure acceleration with developing laser facilities.
To save computational resources we employed a non uniform grid in the transverse direction, i.e. a constant cell spacing is maintained in a region around the axis and then gradually stretched towards the edge. This allows us to keep a high resolution in the center and contain the expanding plasma with a feasible number of grid points. The simulation box is $93\lambda$ wide along $x$ (the laser-propagation direction) and $120\lambda$ along $y$ and $z$. In the central region ($93\times 60\times60\,\lambda$) the cell size is $\Delta x=\lambda/44$, $\Delta y=\Delta z=\lambda/22$.  The grid size is $4096\times1792\times1792$ cells and 64 macro particles per cell per species are used yielding a total number of $\simeq 2\times 10^{10}$. The simulations were run on 16384 BlueGene/Q cores on FERMI at CINECA (Bologna, Italy).
The target is composed of a first layer of ions with charge to mass ratio $Z/A=1/2$ (e.g. C$^{6+}$), $\ell_t=\lambda$ and $n_e=64n_c$ (so that $\zeta=201$), and a second layer of protons, having thickness $\ell_r=\lambda/22$ and density $n_e=8n_c$. 
The laser pulse has amplitude $a_0=198$, a transverse Gaussian profile with waist diameter $w=6\lambda$ and a longitudinal $\cos^2$-like profile with a FWHM duration $\tau_p=9\lambda/c$, all referred to the profile of the fields.
Simulations have been performed both using circular (CP) and linear (LP) polarization.

Fig.\ref{fig_3D} shows a 3D density snapshot for both electrons and ions at intermediate stages of the acceleration process for a simulation with optimal amplitude $a_0=198$ and CP. 
A front view reveals a transverse, net-like structuring of the ion density, which is rather evident in the protons. In the LP case (not shown) in which $a_0=198\sqrt{2}$ there is a tendency of the structures to lengthen along the polarization direction. 

\begin{figure}[t]
\centering
\includegraphics[width=0.5\textwidth]{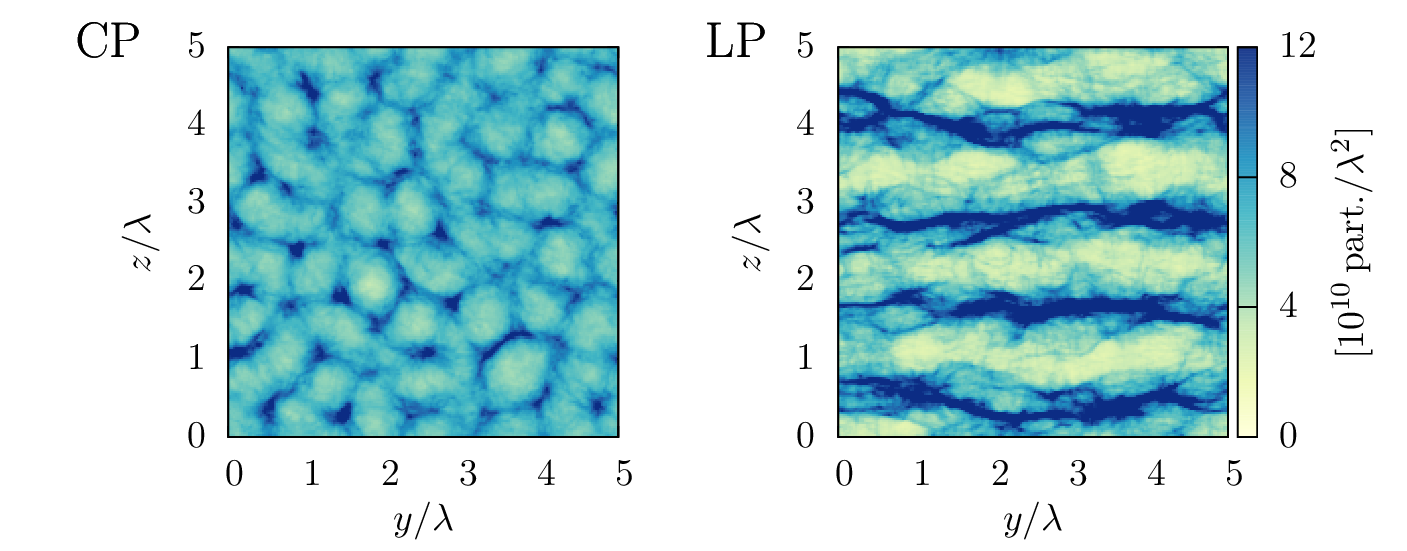}
\caption{(color online) Areal density of carbon ions at $t=15T$ in 3D simulations with same parameters of Fig.\ref{fig_3D} but for a plane wave, for circular (CP) and linear (LP) polarization. 
\label{fig_slab}}
\end{figure}

The difference in the transverse structures between CP and LP is particularly evident for plane wave 3D simulations (performed with {\tt PICCANTE}), where we took an uniform intensity profile and $5\lambda$ as the grid length in $y$ and $z$. In the CP case we observe a pattern of hexagonal-like structures, which indeed corresponds closely to a theoretical prediction, based on symmetry arguments, for a stable structure of the flow in the nonlinear 3D development of the RTI \cite{abarzhiPRE99,*abarzhiPS08}. It is noticeable that such structure provides an example of spontaneous symmetry breaking in a classical system \cite{michelRMP80}, where the continuos  symmetry group of rotations and translations of the initial pulse-target system is reduced to the discrete ``wallpaper'' group p6mm \cite{schattschneiderAMM78}.
For LP, the structures are strongly elongated along the polarization direction, which confirms that the laser electric field ``sweeps out'' the modulations. For both LP and CP, the transverse structures are visible also in the electron density (not shown) and already at $t\simeq10T$ (a faster growth being apparent for LP).

The field modulation and local enhancement due to sub-wavelength surface rippling may play a role in other phenomena related to intense laser interaction with an overdense plasma. As an example we mention the generation of current filaments from the interaction surface, which in several simulations is correlated with a local rippling \cite{sentokuPoP00,mulserLP00}. The transverse modulation of the field may lead to a modulation of the energies for the electrons there accelerated by the ${\bf v}\times{\bf B}$ force, providing a seed for the filamentation instability \cite[and references therein]{califanoPRL06} and explaining why the laser wavelength is the preferred scale for the filaments \cite{sentokuPoP00,lasinskiPoP99,*sentokuPRE02}. 
We also notice that the local transverse flow of momentum ($P_y$) at a rippled surface may lead to the generation of patterns of steady electric and magnetic fields with the ripple periodicity, which could also affect the formation of filaments. 

In conclusion, we showed that self-consistent modulation of radiation pressure and plasmonic enhancement at a rippled surface strongly affect the laser-driven  Rayleigh-Taylor instability, setting a dominant scale close to the laser wavelength as observed in simulations. Three-dimensional simulations show the formation of net-like structures with approximate hexagonal ``wallpaper'' symmetry, in agreement with theoretical predictions.

We thank P.~Londrillo (University of Bologna and INFN, Italy) for help with the \texttt{ALaDyn} code and M.~Lupetti (Ludwig-Maximilians-Universitaet, Muenchen, Germany), F.~Califano (University of Pisa, Italy) and D.~Del Sarto (University of Nancy, France) for useful discussions.
We acknowledge PRACE for access to the BlueGene/Q ``FERMI'', based in Italy at CINECA, via the project ``LSAIL''. Support from MIUR, Italy, via the FIR project ``SULDIS'' is also acknowledged.


\hyphenation{Post-Script Sprin-ger}

\end{document}